\documentstyle[epsf]{article}
\newcommand{\bee}{\begin{equation}}
\newcommand{\ee}{\end{equation}}
\newcommand{\beea}{\begin{eqnarray}}
\newcommand{\eea}{\end{eqnarray}}

\newcommand{\ewxy}[2]{\setlength{\epsfxsize}{#2}\epsfbox[10 60 640 570]{#1}}

\begin{document}
\thispagestyle{empty}
\parskip=12pt
\raggedbottom

\def\mytoday#1{{ } \ifcase\month \or
 January\or February\or March\or April\or May\or June\or
 July\or August\or September\or October\or November\or December\fi
 \space \number\year}
\noindent
\hspace*{9cm} COLO-HEP-407\\
\vspace*{1cm}
\begin{center}
{\LARGE Instanton Content of the $SU(3)$ Vacuum}

\vspace{1cm}

Anna Hasenfratz  and 
Chet Nieter  \\
Department of Physics \\
 University of Colorado,
Boulder CO 80309-390

\vspace{1cm}

\mytoday \\ \vspace*{1cm}

\nopagebreak[4]

\begin{abstract}

We study the topological content of the SU(3) vacuum using 
the renormalization group (RG) mapping method.
RG mapping is a simple smoothing algorithm, 
in which a series of APE-smearing steps are done while 
the topological content of the
configuration is carefully monitored.
This monitoring process makes it possible to separate true
topological objects from vacuum fluctuations and allows an
extrapolation to zero smearing steps.
Using RG mapping  we have measured the instanton distribution
and topological susceptibility for SU(3) gauge theory.  We arrive at
a value for the topological susceptibily, ${\chi}^{1/4}$ of 203(5) MeV.
The size distribution peaks at $\rho=0.3$fm, and is in good agreement with
the prediction of instanton liquid models.

\end{abstract}

\end{center}
\eject


\section{Introduction}

Instantons play an essential  role in the QCD vacuum. In
addition to explaining the U(1) problem \cite{U1,WV}, there is growing evidence
both from phenomenological models \cite{Diakonov,Shuryak_long} and lattice simulations  
\cite{Negele,SU2_SPECTR}  that they play a major role in chiral symmetry
breaking
and the low energy hadron spectrum.  

Phenomenological  instanton liquid models (ILM) describe the
propagation of quarks in the instanton vacuum as hopping from instanton to
instanton. This "hopping propagation" can happen only if the vacuum is filled with
instantons and anti-instantons that overlap and provide a continuous path for the
propagation. To understand if such a path is formed one has to determine the
location and size distribution of instantons in the vacuum.

Over the last few years we have developed several methods, all based on
renormalization group transformations, to study the vacuum of gauge theories
\cite{INSTANTON2,SU2_DENS,RG_SU2}.
Until now we applied these methods to SU(2) gauge  theories. 
Our results for SU(2) gauge  theory indicate that while the
instantons of the vacuum do not form a dense liquid, they do form some kind of
percolating liquid that is sufficient for the hopping propagation
\cite{SU2_DENS,RG_SU2}. 
We also studied the role of topology by extracting the location 
and size of the individual instantons of
the vacuum and built artificial configurations that contained only the
instantons of the vacuum but
none of the vacuum fluctuations or other topological objects \cite{SU2_SPECTR}.
On these smooth instanton 
configurations we found evidence for chiral symmetry breaking supporting the ILM
picture.

This paper is the first in a series where we attempt to understand the role of
instantons in the more realistic SU(3) vacuum.
Here we concentrate on
the properties of the instantons. By generalizing the RG mapping 
method we developed for
the SU(2) model,  we measure the topological charge and also extract the location and
size of individual topological objects of the vacuum configurations. 
The RG mapping method is a  simple and fast
procedure. It is  based on renormalization group methods but at the end
it is only a series of APE smering steps. What distinguishes it from other
algorithms is that we carefully monitor the change of the configurations while doing
as few smearing steps as possible, and extrapolate back to zero smearing. With
that we can minimize the effect of the smearing procedure while use the smoothing
effect to get rid of the vacuum fluctuations.
We find for the topological susceptibility the value $\chi^{1/4}=203(5)$MeV, in
agreement with recently published calculations 
\cite{Deforcrand_SU3,Teper_98,SCRI_susc,Pisa_SU3}. 
However,  our instanton size distribution is quite different form the published
results. We find that the instanton distribution peaks at around $\rho=0.3$fm.
This value is significantly smaller than those of Ref.\
\cite{Deforcrand_SU3,Teper_98} but  in excellent agreement with the ILM
predictions \cite{Shur_dist}.
Our results for 
the density of the instantons in the vacuum is also in agreement with
phenomenological expectations. We found the density to be
about 1.1fm$^{-4}$.

The rest of this paper is organized as follows. In Chapter 2 we will discuss the
RG mapping method and demonstrate why it is necessary to monitor the instantons
during the smearing procedure. Chapter 3 is our results for the topological
susceptibility and instanton size distribution. 
Chapter 4 is the summary.

\section{RG mapping}

Instantons carry only a few percent of the total action of typical Monte Carlo
configurations - they are hidden by the vacuum fluctuations. 
The goal of any method designed to
reveal the vacuum structure is to reduce the short range quantum fluctuations
in the gauge fields while preserving the topological content of the
vacuum.  

The method of inverse blocking \cite{SPINMODEL,INSTANTON1,INSTANTON2} is one of the  
theoretically best supported smoothing algorithms. For any coarse lattice the inverse block
transformation finds the
smoothest 
lattice with half the lattice spacing and twice the lattice size that blocks back
into the original lattice, and consequently has the same topological structure as
the original lattice.
In Ref.\ \cite{INSTANTON2} we have used the inverse blocking method to calculate
the topological susceptibility of the $SU(2)$ vacuum. Unfortunately inverse
blocking  requires 
an extensive amount of computer resources and cannot be used repeatedly due to
rapidly increasing memory requirements. Therefore we could not use inverse blocking to
measure the instanton size distribution of the vacuum or do a reliable scaling
test for the susceptibility by going to smaller lattice spacing and larger
lattices. 

The method of RG cycling \cite{SU2_DENS} combines
the inverse blocking with an additional blocking step that makes it possible
to apply it repeatedly until the gauge configuration
becomes sufficiently smooth. That way it is possible to reveal the instanton size
distribution of the vacuum, but because RG cycling still 
contains an inverse blocking step, it
is also computer intensive and cannot be used on large lattices. 

Our next method, RG mapping, was designed to mimic the
inverse-blocking-blocking sequence of the RG cycling without actually
performing the inverse blocking step. 
In Ref.\ \cite{RG_SU2} the RG cycling transformation 
for $SU(2)$ was fitted to a simple transformation
where each link is replaced by an APE-smeared link\ \cite{APEBlock}
\beea
X_\mu(x) = (1-c)U_\mu(x) & +  & c/6 \sum_{\nu \ne \mu}
(U_\nu(x)U_\mu(x+\hat \nu)U_\mu(x+\hat \nu)^\dagger
\nonumber  \\
& + & U_\nu(x- \hat \nu)^\dagger
 U_\mu(x- \hat \nu)U_\mu(x - \hat \nu +\hat \mu) ),
\label{APE}
\eea
with  $X_\mu(x)$  projected back onto $SU(2)$.  The best fit between a
RG cycled lattice and a smeared lattice was obtained by chosing $c=0.45$ and
performing two smearing steps.  This 
transformation reduces the short range fluctuations effectively and reproduces the
result of RG cycling at least for instantons that are larger than about 1.5
lattice spacing, yet the computer
resources that it uses are insignificant compared to those required
for RG cycling.  

While for SU(2) we adjusted the parameters of the RG mapping method to match one
RG cycling step, the exact correspondence is really not necessary as long as the
evolution of the configuration is monitored.
Nevertheless we decided to use the same smearing parameter $c=0.45$ for the present 
SU(3)
study. We found that with that choice 15-30 smearing steps were necessary to reveal
the vacuum structure of the configurations.

\subsection{The effect of RG mapping on instantons}

During RG mapping, just like during RG cycling, the charge density profile (i.e.
instanton size) changes slowly and usually monotonically. Over several
smearing steps  these changes can be significant requiring a careful monitoring of
the configuration as it is smoothed.  Since we are interested in the the
topological content of 
the original and not the smoothed configurations, an extrapolation to
zero
smoothing steps must be done.
 Fortunately the location of the instantons are fairly
stable and monitoring of the instantons can be done straightforwardly.

\begin{figure}
\centerline{\ewxy{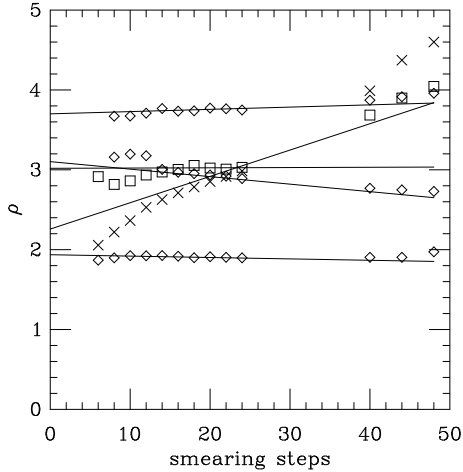}{80mm}}
\caption{Radius versus APE-smearing steps of instantons (for clarity, two symbols,
diamonds and
squares both denote instantons) and anti-instantons (crosses) on a $16^4$ $\beta=6.0$ configuration.}
\label{fig:inst_evol}
\end{figure}

Figure \ref{fig:inst_evol} illustrates how  typical instantons change during the
smearing procedure. This $\beta = 6.0$ ,  $16^4$ configuration has a total
charge of $Q=4$. We followed the evolution of the instantons between 6 and
24 smearings with smearing parameter $c=0.45$, and than again between 40 and 48
steps, to study their behavior after a large
number of smearing steps. 
A total of five stable objects were found on this configuration,
four instantons and one anti-instanton.
The location of each of the identified 
objects was stable, it
 changed by no more than one lattice units in any of the coordinates of the lattice.
  Figure \ref{fig:inst_evol} shows
the sizes of these objects as the configuration is smeared.  The solid lines
are linear extrapolations based on smearing steps 16 to 24.
The slopes of all the 
extrapolations are less than $0.03$, except for the anti-instanton, which changes more
rapidly.   The linear extrapolations match the evolution 
out to large smearing steps for three out of the five objects.  Since the total charge
on this configuration is $Q=4$,  the instanton finding algorithm either missed
an instanton or the anti-instanton is  not there, it is only
a miss-identified vacuum fluctuation. The latter option is supported by the fact
that the anti-instanton changes more rapidly than the instantons of the
configuration.

\section{Results }

\subsection{Parameters of the Simulation}

We have used Wilson action at couplings $\beta=5.85,6.0$ and $6.1$.  The parameters of
the simulations are given in Table \ref{tab:parameters}.  The configurations
were separated by 150 sweeps and all of the lattices are periodic.  The 
lattice spacings in Table \ref{tab:parameters} were calculated using the
results for the $r_0$ parameter and 
the parametrization of the string tension 
of  Ref.\ \cite{Heller}.  Since the string tension in \cite{Heller} is set
at $\sqrt{\sigma}=465$MeV, about 5\% higher than the customary
$\sqrt{\sigma}=440$MeV value, our lattice spacings are somewhat smaller than in
other works.

\begin{table*}[hbt]
\caption{Parameters of the simulations}
\label{tab:parameters}
\begin{tabular*}{\textwidth}{@{\extracolsep{\fill}}|l|c|c|c|c|}
\cline{1-5}
$\beta$  & L  & number of & a [fm] & La [fm]  \\
         &   & configurations &  &          \\
\cline{1-5}
5.85 & 12 & 150 & 0.121(1) & 1.46             \\
\cline{1-5}
6.0 & 12 & 426 & 0.093(1) & 1.12             \\
\cline{1-5}
6.0 & 16 & 100 & 0.093(1) & 1.49             \\
\cline{1-5}
6.1 & 16 & 212 & 0.079(1) & 1.27             \\
\cline{1-5}
\end{tabular*}
\end{table*}

\subsection{The Topological Susceptibility}

We have used the FP algebraic operator developed for SU(2) in \cite{SU2_DENS} to
measure the topological charge. The operator usually gave an integer value within
a few percent after 12 APE smearing steps and remained stable during the
smoothing process. Occasionally it started to change rapidly settling at a
different integer value after a few smearing steps. This change describes the
disappearance of an instanton from the lattice. We have seen that phenomena in
SU(2) as well. The charge for SU(3) was generally more stable under
smoothing than for SU(2). This is probably due to the suppression of very small 
instantons in SU(3) as we will discuss in the next section.

\begin{figure}
\centerline{\ewxy{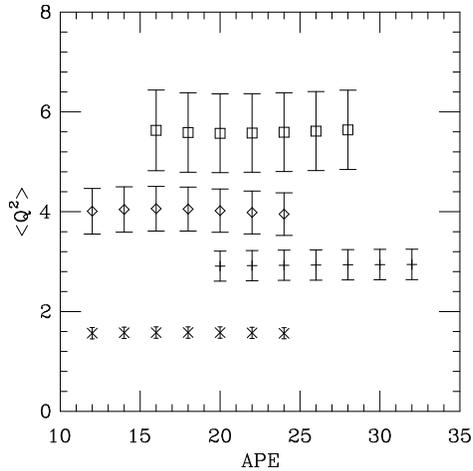}{80mm}}
\caption{The susceptibility $\langle Q^2\rangle$ vs number of
$c=0.45$ APE  steps.  Symbols are diamonds for $\beta=5.85$, crosses for
$\beta=6.0$ on $12^4$ lattices, squares for $\beta=6.0$ on $16^4$
lattices, and pluses for $\beta=6.1$.} 
\label{fig:chi_vs_sweep}
\end{figure}

We have monitored the change of the topological charge under smoothing carefully.
The charge was measured every two APE steps for
each configuration 
between smoothing steps 12 
and  24 for the $\beta = 5.85$ and the $12^4$ $\beta = 
6.0$ configurations, between 16 and 
28 steps for the $16^4$ $\beta = 6.0$ configurations, and
20 and 32 steps for the
$\beta = 6.1$ configurations.  Fig. \ref{fig:chi_vs_sweep} 
shows the average value of the charge squared $<Q^2>$ versus 
the number of smearing steps.  Since the susceptibility is
stable, there is no reason
to extrapolate back to zero smearing steps.  Table \ref{tab:chi_sum}
contains the results for $<Q^2>$ and $\chi^{1/4}$.  The given 
errors in $\chi^{1/4}$ are due to the statistical error of $<Q^2>$.
The errors due to the uncertainty in the lattice spacing are
negligible.

\begin{table*}[hbt]
\caption{Results for the susceptibility }
\label{tab:chi_sum}
\begin{tabular*}{\textwidth}{@{}|l@{\extracolsep{\fill}}|c|c|c|c|}
\hline
$\beta$  & L  & smearing & $\langle Q^2 \rangle$ & $\chi^{1/4}$[MeV]   \\
\hline
5.85 & 12 &                16 & 4.1(4) & 192(5)         \\
     &    &                20 & 4.0(4) & 191(5)         \\
     &    &                24 & 4.0(4) & 191(5)         \\
\hline
6.0 &12  &                16 & 1.6(1) & 198(3)       \\
    &    &                20 & 1.6(1) & 198(3)       \\
    &    &                24 & 1.6(1) & 198(3)       \\
\cline{2-5}
    & 16 &                16 & 5.6(8) & 203(7)       \\
    &    &                20 & 5.6(8) & 203(7)       \\
    &    &                24 & 5.6(8) & 203(7)       \\
\hline
6.1 & 16 &                20 & 2.9(3) & 203(5)        \\
    &    &                26 & 2.9(3) & 203(5)        \\
    &    &                32 & 2.9(3) & 203(5)        \\
\hline
\end{tabular*}
\end{table*}

The susceptibility increases by about 5\% from $\beta=5.85$ to $\beta=6.0$ 
but stabilizes at higher values of $\beta$.  This difference can be atributed to the lack of
small instantons at $\beta=5.85$ due to the larger lattice spacing.
The $\beta=6.0$, $12^4$ lattices likely show a small finite volume effect.
We arrive at a value for the $\chi^{1/4}$ of $203(5)$MeV.  Before we compare
our results with those from other work,
we should first note the string tension value we used is higher than the
standard value by about 5\%. Had we used the customary $\sqrt \sigma =440$MeV value,
we'd obtained  $\chi^{1/4}=192(5)$MeV, in complete
agreement with the results from Ref.\ \cite{Deforcrand_SU3,Teper_98,SCRI_susc}.

\subsection{The Instanton Size Distribution}

\begin{figure}
\centerline{\ewxy{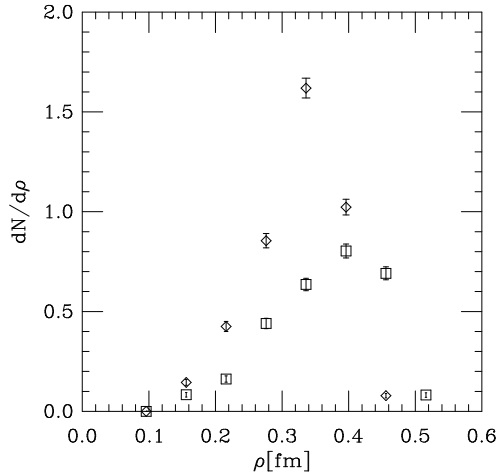}{80mm}}
\caption{The size distribution on 12 (diamonds) and 24 (squares) times smeared $12^4$ lattices
at $\beta=6.0$.}  
\label{fig:step_dist}
\end{figure}

We have identified instantons every 2 APE-smearing steps between 16 and
32 smearing steps with parameter c=0.45. 
 The identification of the instantons and the measurement of
their sizes was done in the same manner as in Ref.\ \cite{SU2_DENS}, 
and the instanton sizes were extrapolated
back to zero smearing steps.
The reason for monitoring instantons is twofold. First, instantons change, usually
grow during the smearing process, as we illustrated in Sect. 2.1. The other reason
is that monitoring helps distinguish true topological objects from vacuum
fluctuations. In Figure \ref{fig:step_dist},
we plot the size distribution of instantons after 12 and 24 smearing steps
at $\beta=6.0$.
The peak of the distribution changes from  0.34fm to 0.40fm as we perform
the additional 12 smearing steps and the shape of the distribution also changes.
The number of objects found decreases by about 50\%,  and
distribution develops a tail at larger $\rho$ values. 
 The development of the tail in the distribution
is consistent
with the observation that the size of the instantons grow with the smearing steps.
The decrease in the number of
objects signals that on the 12 times smeared lattice we identified many
fluctuations as topological objects.
In fact, even after 24 smearing steps the instantons density is still
2.9fm$^{4}$, close to three times larger than the phenomenologically expected
value. That indicates that even after 24 smearing steps we still have a lot of
"spurious instantons". 
To counter this we only identify an object as an instanton
or anti-instanton if it appears to be stable over several smearing steps.
To call an object stable we require not only that its location is stable during
smoothing but also that its size does not change too fast. We approximate the
change of the size linearly and
if the slope of the linear fit
is above some cut-off, the object is rejected.  We choose this cut-off value
so that the charge calculated by taking the difference between the total number of
instantons and anti-instantons found on the configurations give the same
topological susceptibility as the charge measured directly with the FP operator
$<Q = (I-A)^2>= <Q_{FP}^2>$.

On the $12^4$  lattices at $\beta = 6.0$ a  cut-off 0.035 per smearing
steps
gave identical susceptibility with the two methods and predicted the
density of
instantons to be 1.1 per fm$^4$. The results on the other configurations
were
consistent with that but the error of $<Q = (I-A)^2>$ were too large to get
a
precise determination of the cut. However, the  instanton density is much
better
measured, and we found that a cut of 0.035 gave consistently 1.1 per fm$^4$
density. Since the same cut-off value was found for SU(2) as well, in our
analysis we 
used a cut-off  0.035, i.e. we
 kept only those instantons whose size in lattice units
changed not faster than 0.035 per
smearing step.

In their recent work Smith and Teper
\cite{Teper_98} also used APE smearing to smooth the configurations. They chose a
smearing parameter c=0.86, almost twice of ours and smoothed the configurations 23
to 46 times.  We do not have much experience about how the larger $c$ parameter effects
the instantons of the lattice. As a rough estimate one can consider the product
$\sqrt{N_s} c$, where $N_{s}$ denotes the number of smearing
steps,
as a characteristic number of
the amount of smearing. Using that quantity, Ref. \cite{Teper_98} has 3-4
times more smearing than this calculation. 
Figure 9. of \cite{Teper_98} is a similar plot to our Figure \ref{fig:step_dist},
showing the size distribution at
$\beta=6.0$ after 23, 28, 32 and 46 smoothing steps with c=0.86 APE parameter. There
the change due to smearing is less pronounced than in our case, as the average
instanton size changes from 0.49fm to 0.53fm.
It is difficult to compare our numbers directly to those of Ref. 
\cite{Teper_98} since we used a different smearing parameter and 
we did not do the fine tuned filtering of
\cite{Teper_98} at this level.
Nevertheless, it is clear that smoothing effects the instanton size distribution and it is
not surprising that in Ref. \cite{Teper_98} the observed distribution peaks at a larger
$\rho$ value.
Our goal is to correct for the change in instanton size by extrapolating the
results back to zero smearing steps.

\begin{figure}
\centerline{\ewxy{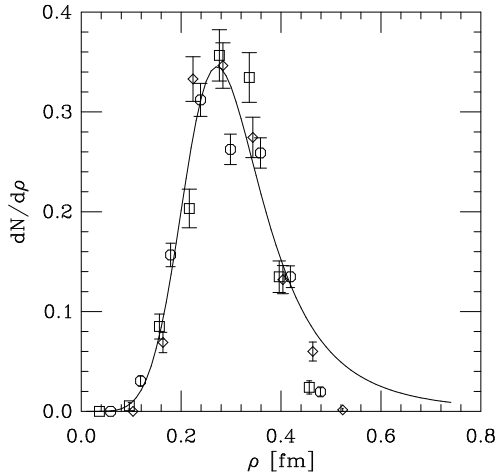}{80mm}}
\caption{The size distribution of the instantons.  The diamonds
correspond to $\beta=5.85$, octagons to $\beta=6.0$, and squares
to $\beta=6.1$.  The first bin of each distribution is contaminated
by the cut-off.  The solid curve is a three parameter fit to the
}
\label{fig:inst_dist}
\end{figure}

Our final result for the instanton size distribution after extrapolation 
is shown in Figure \ref{fig:inst_dist}.  We overlay the data obtained at
$\beta=5.85$ (diamonds), $\beta=6.0$ (octagons), and $\beta=6.1$ (squares).  
The data points for $\beta=6.0$ is a combination of the results from both
lattice sizes. The bin size is 
0.06 fm and the distribution $dN/d\rho$ is measured in fm$^{4}$.  The second
bin of each distribution starts at $\rho=1.6a$.  Because smearing cannot
identify instantons below a certain size, the first bin in each distribution
is unreliable.

The solid line is a fit to a two loop perturbative
instanton distribution formula with a
``regularized'' log \cite{Shur_dist}
\beea
S_I={8\pi^2 \over g^2(\rho)} = b_0L + b_1 \log L ,
\nonumber \\
L={1 \over p}\log[(\rho\Lambda_{inst})^{-p}+C^{p}],
\label{REGLOG}
\eea
where $b_0={11 \over 3} N_c$,  $b_1={17 \over 11} N_c$,
and $p$
and $C$ are arbitrary parameters.
One would expect Eqn. \ref{REGLOG} to be valid if the instanton size distribution is
regulated by quantum fluctuations.
Our fit corresponds  to $p=5.5$, $C=5.5$ and $\Lambda_{inst}=0.71fm^{-1}$. This
$\Lambda_{inst}$ corresponds to $\Lambda_{\bar M\bar S}= 1.45\Lambda_{inst}=206$MeV,
considerably smaller than the value $\Lambda_{\bar M\bar S}=270$MeV obtained from recent
string tension data. The implication of this discrepancy, if any, is not clear.
After all, Eqn. \ref{REGLOG} is only a phenomenological form, even if  the
instanton size distribution is
regulated by quantum fluctuations, there is no proof that the logarithms should be
regularized this way. 

The main difference between the SU(3) distribution and the one we found for SU(2)
\cite{SU2_DENS,RG_SU2} is the lack of small instantons. For small instantons
the instanton distribution can be described by the semiclassical dilute instanton
model which predicts that ${dN \over d\rho} \sim \rho^{b_0}$, therefore small
instantons are indeed suppressed in SU(3) relative to SU(2).
That explains why the topological charge and the topological susceptibility are more
stable under smoothing in SU(3).

The three different coupling values predict a consistent size 
distribution indicating scaling. The
size distribution peaks at around $\bar \rho =0.3$fm.
This value is considerably smaller than  the results of Ref.\ \cite{Deforcrand_SU3}.
who found $\rho_{peak}=0.6$fm or the result of Ref.\ \cite{Teper_98} who quote
$\bar\rho=0.5$fm. The difference between our result and \cite{Teper_98} is 
most likely due to
our extrapolation to zero smearing step. Ref.\ \cite{Deforcrand_SU3} uses
improved cooling as a smoothing method.
 We do not have direct comparison of the techniques. It would be
interesting to monitor individual instantons under cooling and see if and how
their sizes change.
The phenomenological instanton liquid model predicts $\rho_{peak}=0.28-0.33$fm,
depending on the $\Lambda$ parameter used in the calculation
\cite{Shuryak_long,Shur_dist}. That value is
consistent with our result.

\section{Summary}

Using the RG mapping method we calculated the topological susceptibility and the
instanton size distribution for SU(3) gauge theory. 
This technique, developed originally for the SU(2) gauge model and based on different
renormalization group methods, is a fast and simple smoothing algorithm. It
consists of several APE smearing steps. What distinguishes it from similar methods
is that only the minimal number of smearing steps are performed and the
topological content of the configuration is continuously monitored. That makes it
possible to separate vacuum fluctuations from true topological objects and to
control the change in size of the instantons during the smearing process.

We found scaling for both the  topological susceptibility and the
instanton size distribution
in the lattice spacing range of  0.12fm to 0.079fm. Our value
for the topological susceptibility, $\chi^{1/4}=203(5)$MeV is consistent with
other recent calculation but our instanton size distribution predicts
considerably smaller instantons. We attribute this discrepancy to the change of
the size of topological objects, which we carefully monitor and remove. Our result
for the instanton size distribution is consistent with the phenomenologically
successful instanton liquid model prediction.
 
\section*{Acknowledgements}
We would like to thank T. DeGrand and T. Kov\`acs for useful conversations, and E.
Shuryak for clarifying many questions regarding the instanton liquid model.
We would like to thank  the Colorado High Energy experimental
groups for allowing us to use their work stations.
This work was supported by the U.S. Department of Energy
 grant DE--FG03--95ER--40894.

\newcommand{\PL}[3]{{Phys. Lett.} {\bf #1} {(19#2)} #3}
\newcommand{\PR}[3]{{Phys. Rev.} {\bf #1} {(19#2)}  #3}
\newcommand{\NP}[3]{{Nucl. Phys.} {\bf #1} {(19#2)} #3}
\newcommand{\PRL}[3]{{Phys. Rev. Lett.} {\bf #1} {(19#2)} #3}
\newcommand{\PREPC}[3]{{Phys. Rep.} {\bf #1} {(19#2)}  #3}
\newcommand{\ZPHYS}[3]{{Z. Phys.} {\bf #1} {(19#2)} #3}
\newcommand{\ANN}[3]{{Ann. Phys. (N.Y.)} {\bf #1} {(19#2)} #3}
\newcommand{\HELV}[3]{{Helv. Phys. Acta} {\bf #1} {(19#2)} #3}
\newcommand{\NC}[3]{{Nuovo Cim.} {\bf #1} {(19#2)} #3}
\newcommand{\CMP}[3]{{Comm. Math. Phys.} {\bf #1} {(19#2)} #3}
\newcommand{\REVMP}[3]{{Rev. Mod. Phys.} {\bf #1} {(19#2)} #3}
\newcommand{\ADD}[3]{{\hspace{.1truecm}}{\bf #1} {(19#2)} #3}
\newcommand{\PA}[3] {{Physica} {\bf #1} {(19#2)} #3}
\newcommand{\JE}[3] {{JETP} {\bf #1} {(19#2)} #3}
\newcommand{\FS}[3] {{Nucl. Phys.} {\bf #1}{[FS#2]} {(19#2)} #3}

\end{document}